# Molecular Valves for Controlling Gas Phase Transport Made from Discrete Angstrom-Sized Pores in Graphene

Luda Wang[1,7], Lee W. Drahushuk[2], Lauren Cantley[3], Steven P. Koenig[4,5], Xinghui Liu[1], John Pellegrino[1], Michael S. Strano[2], and J. Scott Bunch[3,6*]

[1] University of Colorado, Department of Mechanical Engineering, Boulder, CO 80309 USA

[2] Massachusetts Institute of Technology, Department of Chemical Engineering, Cambridge, MA 02139 USA

[3] Boston University, Department of Mechanical Engineering, Boston, MA 02215 USA

[4] National University of Singapore, Department of Physics, 117542 Singapore

[5] National University of Singapore, Center for Advance 2D Materials and Graphene Research Centre, 117546 Singapore

[6] Boston University, Division of Materials Science and Engineering, Brookline, MA 02446 USA

[7] Massachusetts Institute of Technology, Department of Mechanical Engineering, Cambridge, MA 02139 USA (current address)

*e-mail: bunch@bu.edu

An ability to precisely regulate the quantity and location of molecular flux is of value in applications such as nanoscale 3D printing, catalysis, and sensor design[1-4]. **Barrier materials containing pores with molecular dimensions have previously been used to manipulate molecular compositions in the gas phase, but have so far been** **unable to offer controlled gas transport through individual pores[5-18]. Here, we show that gas flux through discrete angstrom-sized pores in monolayer graphene can be detected and then controlled using nanometer-sized gold clusters, which are formed on the surface of the graphene and can migrate and partially block a pore. In samples without gold clusters, we observe stochastic switching of the magnitude of the gas permeance, which we attribute to molecular rearrangements of the pore. Our molecular valves could be used, for example, to develop unique approaches to molecular synthesis that are based on the controllable switching of a molecular gas flux, reminiscent of ion channels in biological cell membranes and solid state nanopores[19].**

We studied 2 types of angstrom pore molecular valves: a porous single layer of suspended graphene with no gold nanoclusters on its surface (PSLG) and a porous single layer of suspended graphene on top of which we evaporated gold nanoclusters (PSLG-AuNCs). To fabricate both types of devices, we start with suspended pristine monolayer graphene which is impermeable to all gases[20] and defect free[21]. The graphene is mechanically exfoliated over predefined etched wells in a silicon substrate with 90 nm of thermal silicon oxide on top. This forms a graphene-sealed microcavity which confines a $\sim\mu m^3$ volume of gas underneath the suspended graphene. We use 2 techniques to introduce molecular-sized pores. The first method uses a voltage pulse applied by a metallized AFM tip[22]. Figure 1a illustrates the method with a ~300 nm diameter pore created in the centre of a graphene membrane by applying a voltage pulse of -5V for 100 ms.

A pressurized blister test is used to determine the leak rate out of the graphene sealed microcavity[23]. The microcavity is filled with pure $H_2$ or $N_2$ at 300-400 kPa and the graphene is bulged up due to the pressure difference across it. An example for an unetched pristine sample pressurized with $N_2$ is shown in Figure 1b. In this instance, after a voltage pulse of -9 V for 2 s to the centre of the membrane a single pore is created—we found that the voltage and time needed to introduce a pore varied depending on the AFM tip used, thus the difference in sizes between Fig. 1a and 1c. Immediately after a pore is formed, the deflection drops and the graphene is flat except for a few wrinkles introduced by the process (Fig. 1c). The AFM image shows no detectable pore meaning that the pore is smaller than the resolution of the AFM. For the PSLG-AuNCs samples, gold atoms are evaporated onto the graphene. Figure 1d shows the graphene sample in Fig. 1c after gold evaporation and repressurization with $N_2$ gas. The graphene is bulged up indicating that the gold clusters blocked the pore.

In the second poration method, a molecular-sized pore is introduced at a random spot on a $H_2$-pressurized, graphene sealed microcavity using UV-induced oxidation[23,24]. We monitor the deflection of the graphene using an AFM operating in intermittent contact mode. Figures 1e-1g shows the maximum deflection vs. time for 3 graphene sealed microcavities labelled "Membrane 1" (left), "Membrane 2" (middle), and "Membrane 3" (right) formed by the same graphene flake covering 3 wells. After exposing all of the membranes to a series of eight etches, "Membrane 2" shows a dramatic change in the deflection just after the last etch indicative of a rapid leaking of all of the $H_2$ gas inside the microcavity, while any counter-transport of air back into the microcavity is much slower. "Membrane 1" and "Membrane 3" have leak rates that are relatively unchanged, suggesting that they remain unporated and pristine. Thus, we see that the initiation of Å-sized porosity in graphene flakes by UV exposure is a sparse and discrete process. The instantaneous change in the permeance from a very slow leak to a very fast leak is strongly indicative of a single pore being formed. It would be highly unusual (though not impossible) for multiple similar size pores to be introduced into "Membrane 2" at once while the other two membranes from the same graphene flake remain unporated.

In the first few hours after UV-induced oxidation of a graphene sample with evaporated gold clusters, the gold clusters are observed to migrate and congregate on the surface of the graphene sometimes blocking the created pore (Fig. 1j-1k). For such a blocked PSLG-AuNC device, the leak rate of $H_2$ gas out of the microcavity is slow initially and the graphene deflection changes slowly with time (Fig. 1o). A fit to a membrane mechanics model (see Supplementary Material) is shown as a red line in Fig. 1s. During scanning there is a sudden jump in the deflection and a much faster leak rate is observed (Fig. 1p and q). This change from bulged up to bulged down occurs in ~30 seconds. A line cut through the centre of the graphene during this process is shown in Figure 1o. From the line cut (blue lines in Fig.1o to q), we can see the maximum deflection drops from 147 nm to -115 nm in ~1 minute (Fig.1s). We attribute this sudden change in the deflection to migration of the gold nanoclusters on the surface of the graphene that "opened" a single Å-sized pore in the graphene allowing a fast leak rate (shown schematically in Fig. 1l to 1n). As the case for "Membrane 2" in Fig. 1f, a single pore is likely responsible for the sudden change in permeance since it is unlikely that multiple pores opened simultaneously.

The change in permeance can be trigged by laser induced heating in vacuum to stimulate movement of the AuNCs (Fig. 2). In this case, the graphene in Figure 1 displays a fast permeance as evidenced by a decrease in the maximum deflection vs. time of $H_2$ gas, taking place in less than 30 minutes. The permeance is relatively constant over many measurements at different starting internal pressures of hydrogen suggesting that eventually the AuNCs have stopped migrating. This is further confirmed by AFM images (Fig. 2a inset left) which show the configuration of AuNCs as stable. After shining a laser on the surface of the graphene, the permeance slows considerably, now taking ~ 30 – 90 minutes to leak out depending on the initial internal pressure. Again the permeance is stable during these measurements over multiple internal pressures. An AFM image of the surface of the graphene after laser exposure shows a change in the configuration of AuNC on the surface of the graphene (Fig. 2a inset right).

From the max deflection vs. time curves (Fig. 2a and 2b), the rate of change of the number of molecules, n, constitutes a leak rate dn/dt that can be extracted assuming a simple membrane mechanics model (Fig. 2c) (see Supplementary Information)[23]. We observe that the leak rate shows a linear dependence on the pressure difference with a slope of 8.41±0.26 x $10^{-24}$ mol-$s^{-1}$-$Pa^{-1}$ before and 1.70±0.02 x $10^{-24}$ mol-$s^{-1}$-$Pa^{-1}$ after laser exposure, consistent with Å-sized pore transport[25]. A histogram of the leak rate normalized by the pressure difference, which we define as the permeance, is shown in Fig. 2d. Counts on the histogram correspond to a permeance calculation based on the slope around each deflection data point. There are clearly 2 defined states of the permeance before and after laser exposure. Additional heating with the laser leads to pore opening showing that this process is repeatable and reversible (see Supplementary

Information) and demonstrates the ability to control the gas flux through a single Å-sized pore molecular valve.

Gas transport through the Å-sized pore can be modelled using classical effusion. When the pore size is smaller than the mean free path of the molecule, classical effusion dictates that the time constant for the decay of the number of molecules in the graphene sealed microcavity is given by:

$$\tau = \frac{V}{\gamma}\sqrt{\frac{2\pi M_w}{RT}} \qquad (1)$$

where $V$ is the volume of the container, $\gamma$ is the transmission coefficient, $M_w$ is the molecular mass, R is the ideal gas constant, and $T$ is the temperature[26]. The transmission coefficient, $\gamma$, incorporates both the physical geometry of the pore, and any energy barrier from molecular interactions between the molecule and the pore. Due to the volume of gas in the graphene sealed microcavity (~1 µm³), the time constant for the effusion due to a single Å-sized can be minutes making our geometry ideally suited for measuring the leak rate through a single sub-nm pore and for observing sub-Å² changes in the transmission coefficient. Correspondingly, the leak rate dn/dt assuming classical effusion is given by:

$$\frac{dn}{dt} = \frac{\gamma}{\sqrt{2\pi M_w RT}} \cdot \Delta p \qquad (2)$$

where $\Delta p$ is the partial pressure difference across the graphene[25]. A plot of dn/dt vs. $\Delta p$ shows a linear dependence further supporting classical effusion (Fig. 2c). The transmission coefficient can be deduced from the slope and changes from 0.0047 Å² (before) to 0.00095 Å² (after) laser heating. The transmission coefficient is the geometric area of the pore (a few Å²) multiplied by a transmission probability that an impinging gas atom or molecule has sufficient energy to pass through the potential barrier of the pore. Hence, $\gamma$ is considerably smaller than the cross sectional area of the test gas $H_2$ (2.89 Å) providing evidence that the pore is on the order of the kinetic diameter of $H_2$, and we are measuring small changes in the energy barrier from molecular interactions between the gas and pore mouth.

The ability to observe small changes in $\gamma$ allows us to vary the molecular size and study how that influences gas transport in the same PSLG. For this study, we used a single PSLG that contained no AuNCs, and four additional etching exposures were done after the first observed rapid change in deflection which indicates a pore formation. The leak rate, dn/dt, of He gas through PSLG shows a linear dependence on $\Delta p$ and is relatively constant over a range of $\Delta p$ from ~ 100 kPa to 700 kPa (Fig. 3a inset). This agrees with classical effusion with a slope equal to 1.5±0.01 x $10^{-23}$ mol·s$^{-1}$·Pa$^{-1}$ corresponding to $\gamma = 0.011$ Å$^2$ (Fig. 3a). The permeance for the other noble gases, Ar, Ne, and He, are shown as a histogram with average values and standard deviations of 3.4 ± 2.1, 25 ± 16, and 145 ± 22 x $10^{-25}$ mol·s$^{-1}$·Pa$^{-1}$, respectively. This follows a trend of a lower permeance for larger gas atoms. In addition to the noble gases, we measured the permeance of the non-noble gas molecules, $H_2$, $CO_2$, and $N_2O$ both before and after introducing the molecular-sized pore (Fig. 3c). Data for the non-porated graphene was taken on 4 separate but similar monolayer graphene membranes and measured leak rates agree well with a leak primarily through the underlying silicon oxide substrate[20,23,27]. In all cases, there was a considerable increase in the leak rate after poration, supporting the conclusion that the leak rate is primarily permeation through the molecular-sized pore.

Using equation 2, we can deduce the transmission coefficient as a function of the kinetic diameter for all the gases measured. As expected for the noble gases, the transmission coefficient increases as the kinetic diameter decreases: (Ar) $0.00084 \pm 0.00053$ Å$^2$, (Ne) $0.0045 \pm 0.0028$ Å$^2$ and (He) $0.011 \pm 0.002$ Å$^2$ (Fig.3d) further confirming that the pore size is on the order of the kinetic diameter and showing the influence of the molecular size on $\gamma$. In addition, the leak rate of $H_2$ roughly follows the trend observed with the noble gas atoms. However, $CO_2$ and $N_2O$ follow a very different trend (inset Fig. 3d). Their transmission coefficients are considerably larger than one would expect from their kinetic diameter ($CO_2$) $0.021 \pm 0.016$ Å$^2$ and ($N_2O$) $0.048 \pm 0.038$ Å$^2$. We attribute this to chemical interactions that $N_2O$ and $CO_2$ have with the pore which lowers the energy barrier for transport. These experiments demonstrate that gas transport of gas molecules with polar bonds clearly shows a strong influence on chemical interactions between the molecule and the pore consistent with recent theoretical calculations[28,29].

Though the permeances are nominally constant over long time periods (days) of measurements, the transmission coefficient demonstrates discretized fluctuations indicative of stochastic switching. Fig. 4a illustrates the concatenated permeances of Ne over time along with a fit to discrete states using Hidden Markov modelling (see Supplementary Material). The data in Fig. 4a was taken over the course of five days where a vertical dashed line corresponds to the start of a new measurement. All measurements are concatenated into a single observed time axis so that the repetition of certain states and values of permeance can be seen. This switching is clearly seen in Fig. 4b where the permeance switches five times within 1 hour (black circles). The histogram of the permeance for Ne is plotted in Fig. 4c. The permeance shows a large number of states on the low end of the spectrum with occasional switching to faster leak rates. We fit these permeance values versus time to discrete states and applied a Hidden Markov model (HMM) to elucidate transition rates[30,31]. We use additional analysis to show that the numerous observed states are consistent with a system having three independent pores with two-states each; the frequency of switching between the states, averaged across three pores, yields an approximate value for the activation energy of the switching process, 1 eV with three two-state pores (See supplementary information). This is comparable to the activation energy required for rearrangement of molecular bonds, such as cis-trans isomerization[32]. These calculations using our experimental results demonstrate that relatively minor changes in a pore's configuration can have an observable impact on its permeation characteristics.

In conclusion, we demonstrated a type of molecular valve in graphene which allowed us to control the gas flux through discrete Å-sized pore. The process was controlled by movement of AuNC on the graphene surface. These results lead to a greater mechanistic understanding of molecular gas transport through molecular-sized pores in atomically-thin materials. The switching observed may lead to unique sensors based on the reversible switching of molecular transport through ~ Å-sized pores reminiscent of ion channels in biological cell membranes.

**Methods**

Device fabrication and pressurization follows closely references[20,23,27,33]. Suspended graphene is fabricated using mechanical exfoliation on silicon oxide substrates with predefined etched wells. The wells were defined by photolithography on an oxidized silicon wafer with 90 nm silicon oxide on top and have a diameter of ~3-5 μm. Reactive

Ion Etching was used to etch the wells to a depth of 400~1000 nm, and the "scotch tape" method was used to deposit graphene over the wells. The gold atoms were evaporated onto the graphene in vacuum (CVC 3-boat thermal evaporator, at 0.1 Å/s for less than 0.5 s). For some devices, we evaporated gold atoms prior to poration and for others we evaporated gold atoms after poration. No significant difference between these was found.

To pressurize the inside of the microcavity, we put the sample into a high pressure chamber with a certain gas species at a prescribed pressure which we call the charging pressure. After several hours or days, depending on the gas species used, the pressure in the microcavity comes to equilibrium with the charging pressure. The pristine graphene sheet is impermeable to any gases, but the gas can diffuse through the silicon oxide substrate. To reach ambient pressure equilibrium, we may wait 4~30 days depending on the gas species.

After measuring the leak rate for the pristine graphene, we etched pores by exposing the graphene to a UV lamp ($\lambda_1$=185 nm, $\lambda_2$=254 nm) under ambient conditions. We first pressurized them with pure $H_2$ up to 200 kPa above ambient pressure. After the microcavity reached equilibrium we removed it from the pressure chamber and measured the deflection using AFM. We continue with a series of short UV exposures followed by AFM scans. Once a pore is created whose size is between $H_2$ and $N_2$, then the deflection abruptly changes from positive to negative. After a pore is created, the leak rate is dominated by transport through the pores and 1~12 h of pressurization, depending on the gas species, is sufficient for equilibration.

To create pores by the voltage pulse method we first pressurize the graphene with $N_2$ to 300 kPa above ambient pressure. We then apply a triggered force curve at ambient conditions using a metallized AFM tip. The bias voltage is applied while the tip is in contact with the surface of the graphene membrane. This is repeated with an increasing magnitude of voltage and contact time until a pore is created. The pore is detected by observing an abrupt decrease in the deflection.

Laser heating of the suspended graphene was accomplished using a He-Ne laser or solid state laser ($\lambda$ = 633 or 532 nm) with a power of ~1 - 10 mW as measured before it enters the vacuum chamber through a sapphire window. The laser spot size is estimated at 3.5 μm with an exposure time of 5 min.

**Acknowledgements**

We thank Xiaobo Yin for useful discussions, and Anna Swan, Bennett Goldberg and Jason Christopher for assistance with Raman spectroscopy. This work was supported by NSF Grants #1054406(CMMI: CAREER: Atomic Scale Defect Engineering in Graphene Membranes), the National Science Foundation (NSF) Industry/University Cooperative Research Center for Membrane Science, Engineering and Technology (MAST), and in part by the NNIN and the National Science Foundation under Grant No. ECS-0335765.

**Author contributions**

L.W., L.C. and S.P.K. performed the experiments. L.W. and J.S.B. conceived and designed the experiments. L.W.D. and M.S.S. developed the theory and modelling. L.W., L.C., S.P.K. and X.L. prepared and fabricated the samples. L.W., L.W.D., J.P., M.S.S. and J.S.B. interpreted the results and co-wrote the manuscript.

**Additional information**

Supplementary information is available in the online version of the paper. Reprints and permission information is available online at http://www.nature.com/reprints. Correspondence and requests for materials should be addressed to J.S.B (bunch@bu.edu).

## Competing financial interests

The authors declare no competing financial interests.

# Figure Legends

**Fig. 1 Fabrication of molecular valves in suspended graphene (a-d) AFM voltage pulse etching to introduce pores in suspended graphene** (a) AFM height image of suspended graphene with a large diameter pore (~300 nm) etched at its centre using AFM voltage pulse etching; (b-c) AFM height image of pressurized suspended graphene before (b) and ~3 min. after (c) etching a small pore (~1 nm) at the centre of the membrane using AFM voltage pulse etching. The pore is below the resolution of the AFM and not visible in the image; (d) The same membrane in (b) and (c) after gold evaporation and pressurization with $N_2$ gas. The bulged up nature shows that it now holds gas; **(e-s) UV etching to introduce pores on suspended graphene** (e-g) Max. Deflection vs. time

during UV etching for three graphene sealed microcavities on the same flake. The UV etching was stopped after Etch 8. (h-i) the AFM height images corresponding to the first and last points in (e-g). (j-k) AFM amplitude images showing the movement of gold nanoparticles on suspended graphene; (l-n) Schematic of the gold nanoparticles (yellow solid circles) blocking and unblocking the pore on the monolayer graphene; (o-q) AFM height images capturing the deflection changes illustrated in (l-n); (r) Deflection vs. position through the centre of the suspended graphene in (o-q); (s) Maximum deflection vs. time for the dramatic leak rate change. The solid red line is a fit to the data before switching using a membrane mechanics model.

**Fig. 2 Controlling the leak rate by laser induced heating.** (a,b) Maximum deflection of the graphene before (a) and after (b) focusing a laser beam at the centre of the graphene. Different colours represent different charging pressures. For (a), the charging pressure sequence is 200 to 700 to 200 kPa, in 100 kPa increments; for (b), the charging pressure sequence is 200 to 850 kPa, in 50, 100, or 150 kPa increments. (inset in a) AFM amplitude images of the suspended graphene corresponding to the state of the graphene deflection for the measurements in (a – left inset) and (b – right inset) (c) Leak rate dn/dt vs. pressure difference $\Delta p$ for (a) - shown in red and (b) – shown in black; (d) Histogram of the permeance from (data in (a) –red) and (data in (b) - black).

**Fig. 3 Leak rates of gases through a porous monolayer suspended graphene without gold nanoparticles** (a) Leak rate dn/dt vs. pressure difference $\Delta p$ for He gas (inset maximum deflection vs. t for He gas) (b) Histogram of permeance for the noble gases, Ar, Ne, and He (c) Permeance vs. kinetic diameter for all of the measured gases before (blue) and after (black) etching. Error bars represent ±s.d. for different measurements on the same membrane. (d) Transmission coefficient (calculated from eq. 2) vs. kinetic diameter for He, Ne, $H_2$, and Ar. (inset) Transmission coefficient (calculated from eq. 2) vs. kinetic diameter for He, Ne, $H_2$, $N_2O$, $CO_2$, and Ar. Error bars represent ±s.d. for different measurements on the same membrane.

**Fig. 4 Stochastic switching of the leak rate through porous monolayer graphene without gold nanoparticles** (a) Permeance (black circles) and fit (red line) vs. time for all the Ne data. Bottom axis, observable time, corresponds to the 800 minutes of measurements taken over five days after repeated pressurization. Each measurement is separated by a dashed line. (b) Single experimental run within (a) matching highlighted time range. Left axis, blue squares - maximum deflection versus time for Ne. Right axis– permeance vs. time calculated from the change in deflection vs. time. (c) Histogram of the permeance (on vertical axis of a) for all the data in (a).

# Figures

## Figure 1

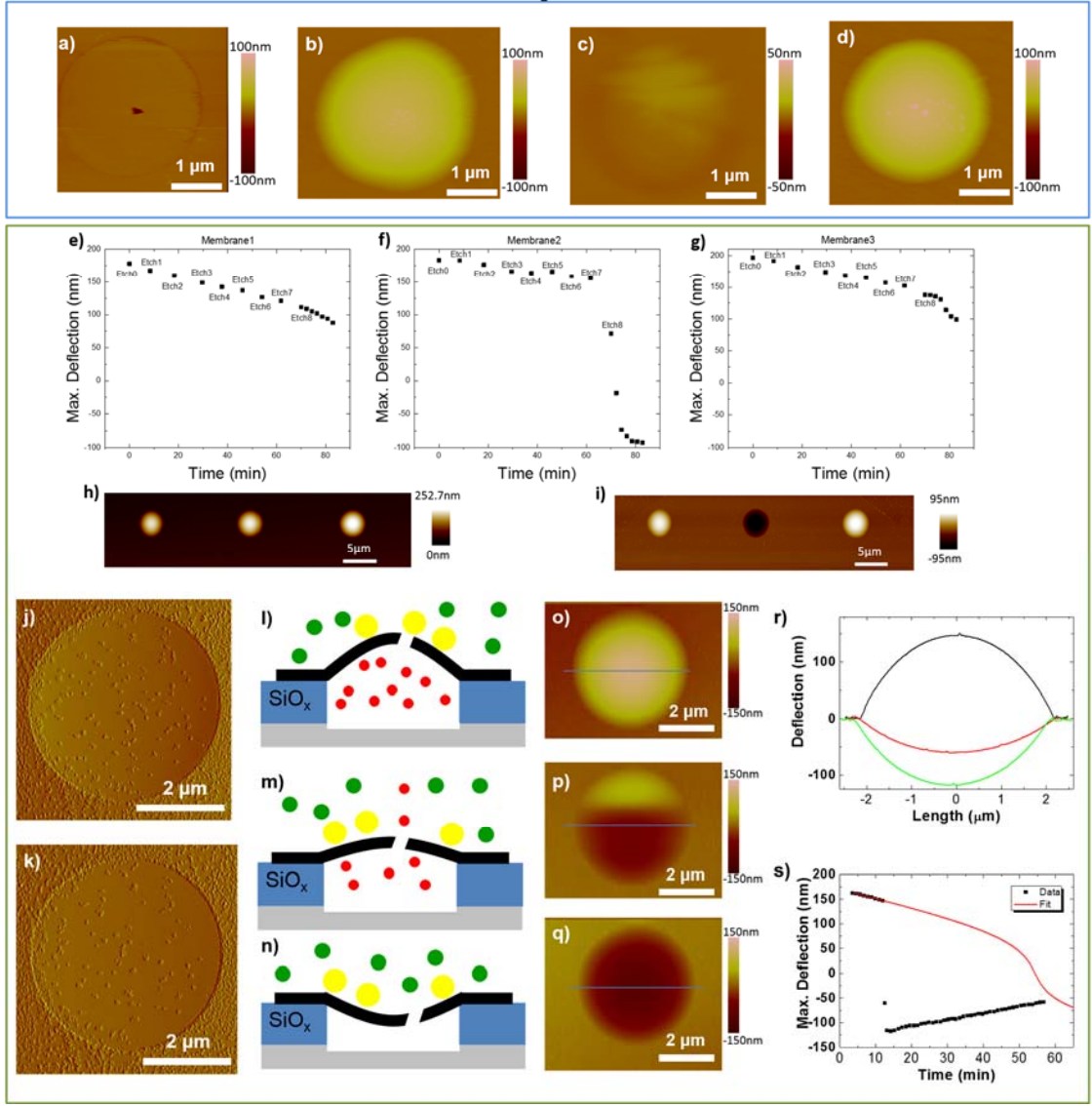

**Figure 2**

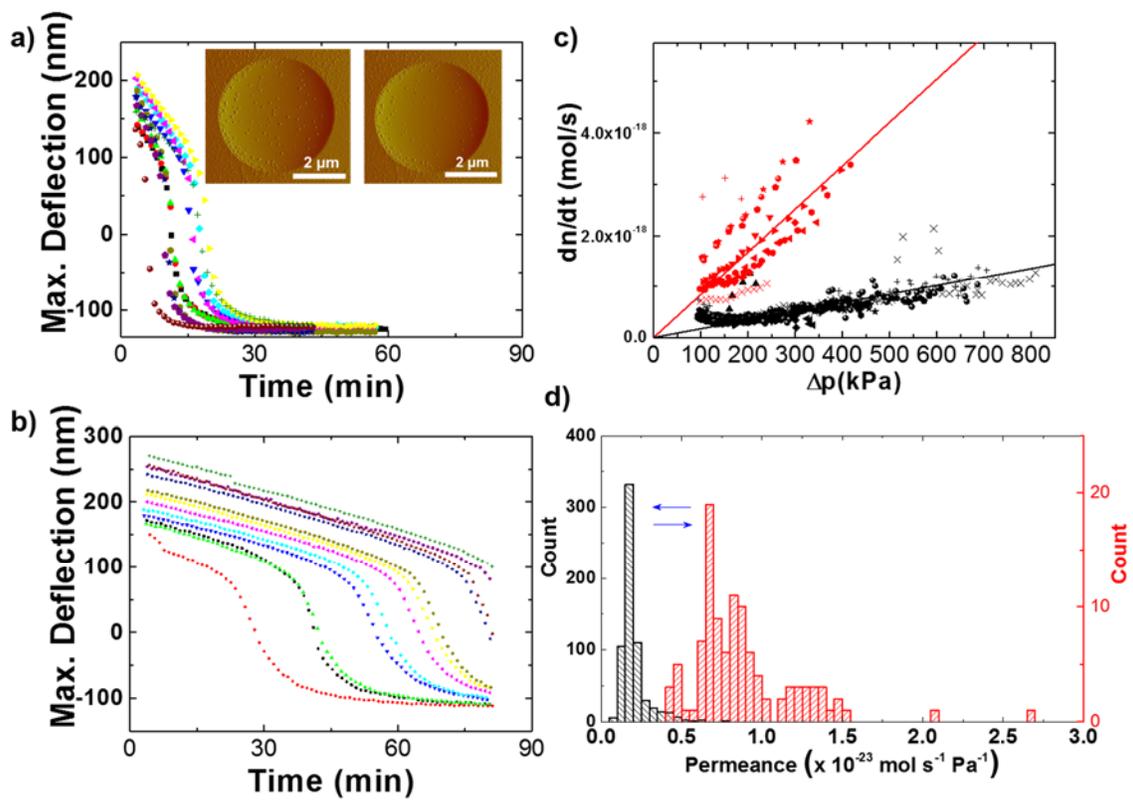

**Figure 3**

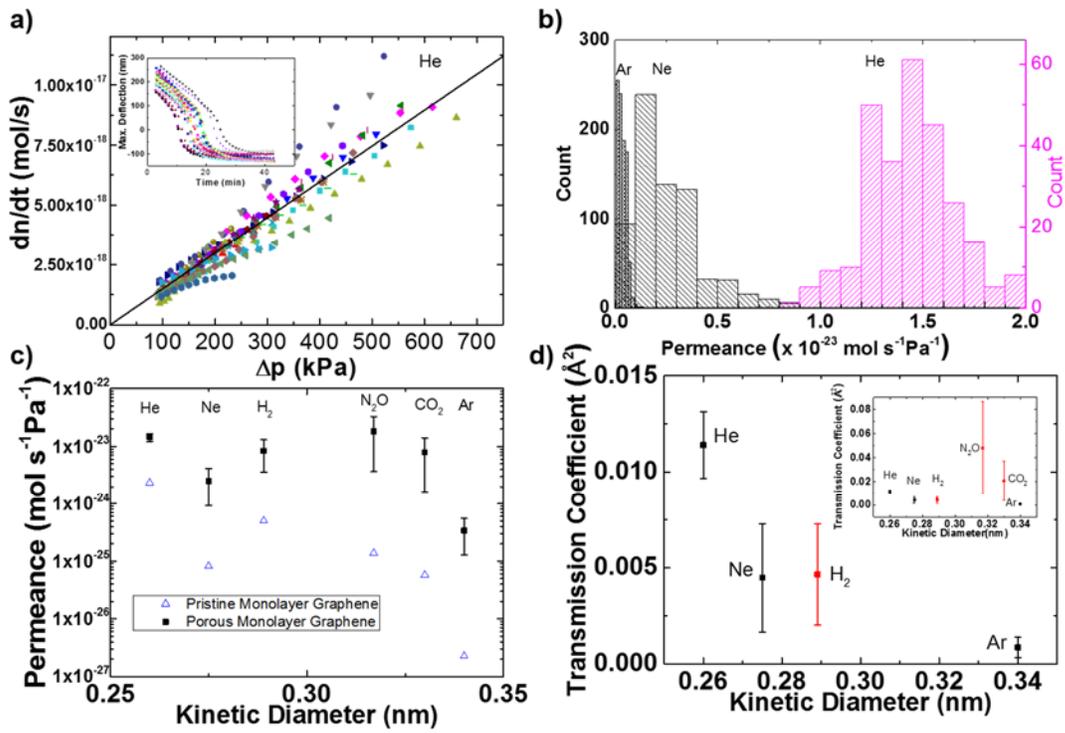

**Figure 4**

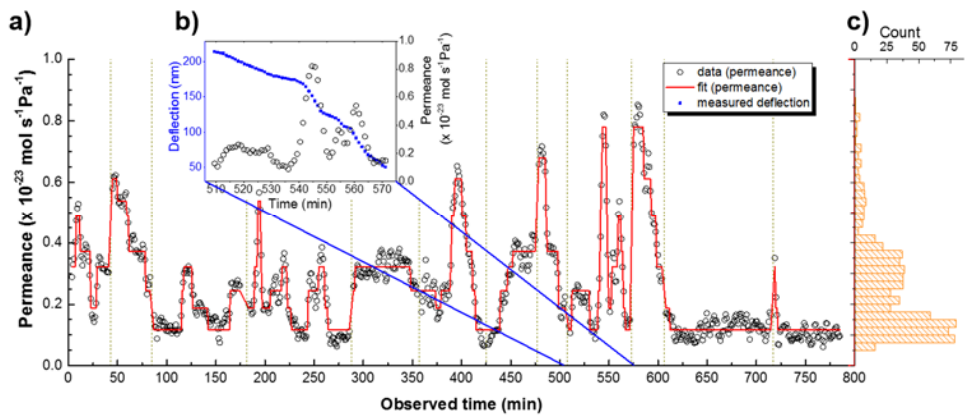

# Molecular Valves for Controlling Gas Phase Transport Made from Discrete Angstrom-Sized Pores in Graphene

Luda Wang[1,7], Lee W. Drahushuk[2], Lauren Cantley[3], Steven P. Koenig[4,5], Xinghui Liu[1], John Pellegrino[1], Michael S. Strano[2], and J. Scott Bunch[3,6*]

*e-mail: bunch@bu.edu

**Supplementary Information:**

1. Leak rate through the silicon oxide substrate

The leak rate can be derived from the ideal gas law and Hencky's solution for a clamped circular membrane and follows closely reference[1]:

$$\frac{dn}{dt} = \frac{1}{RT}[3K(v)\frac{Et}{a^4}\delta^2 \cdot (V_0 + V_b) + \left(P_{atm} + K(v)\frac{Et}{a^4}\delta^3\right) \cdot C(v)\pi a^2] \cdot \frac{d\delta}{dt} \qquad (S1)$$

where $R$ is the gas constant, $T$ is the temperature, $E$ = 1 TPa is the Young's modulus, $t$ = 0.335 nm is the thickness of the membrane, $a$ is the radius of the membrane, $V_0$ is the microcavity volume at zero deflection, $V_b$ is the bulged up volume, $P_{atm}$ is the ambient pressure, $\delta$ is the maximum deflection of the membrane, $K(v = 0.16) = 3.09$, and $C(v = 0.16) = 0.52$ are constants determined by the Hencky's solution.

The maximum deflection versus time was measured for pristine unetched graphene (Supplementary Fig. 1). The samples were inserted into the high pressure chamber with ~300 kPa charging $H_2$. After a few weeks, the internal pressure of the microcavity reached equilibrium with the charging pressure. Continuous AFM scanning was taken during the first 100 min of removal from the pressure chamber, and the deflection decreased by a few nanometres. From the slope of the deflection vs. time and equation S1, we determine that the permeance is ~6 x $10^{-25}$ mol s$^{-1}$Pa$^{-1}$, which is at least one order of magnitude lower than the $H_2$ permeance of the porous graphene.

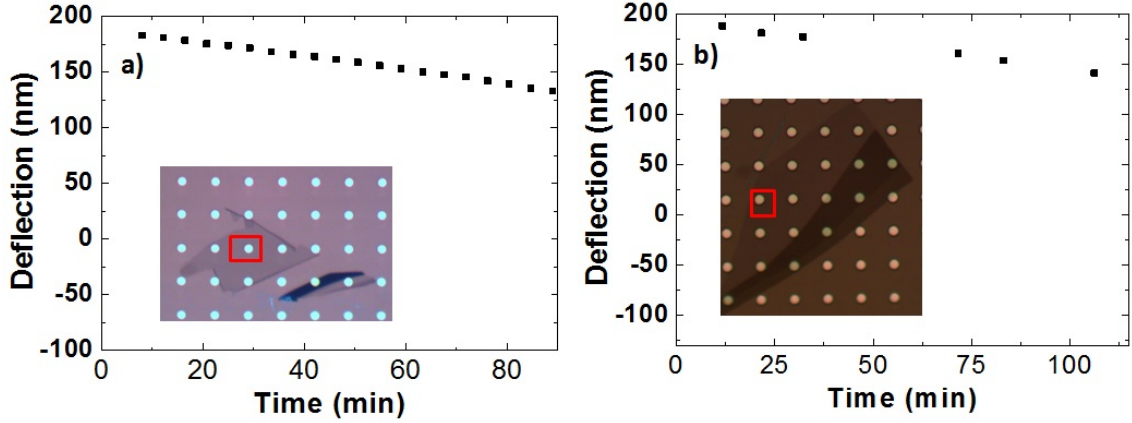

**Supplementary Fig. 1: Maximum Deflection vs time before etching** (a) Maximum deflection vs time for the PSLG-AuNC in Figure 1 and 2 before poration; (insets): optical image of the monolayer graphene flake; (b) Maximum deflection vs time for the PSLG in Figure 3 and 4 before poration (inlay): optical image of the monolayer graphene flake;

2. Description of data analysis (calculating permeance vs time)

Due to the stochastic nature of the switching of transport in the porous graphene, we developed a solution to determine the permeance as a function of maximum deflection. This is different than the solution of equation S1 from which we extract a constant permeance. For the porous graphene, we first used least square fit method to smooth the dn/dt data. In order to fit the data, we developed a model that describes the deflection of the microcavity versus time. We start with a derivation based on the ideal gas law with $P$ (pressure), $V$ (volume), $n$ (mols), and T (temperature) describing the state of the microcavity. The differential with respect to time describes the rate at which molecules leave the chamber, which is the transport rate.

$$PV = nRT \tag{S2}$$

$$\frac{d}{dt}(PV) = RT\frac{dn}{dt} \tag{S3}$$

Pressure and deflection are related by the mechanical properties of the graphene and geometry, therefore pressure ($P$) can be described as a function of deflection. Similarly, the volume of the well is directly related to the deflection. Therefore the ideal gas law can be written with those two terms as function of the deflection.

$$\frac{d\delta}{dt}\frac{d(P(\delta)V(\delta))}{d\delta} = RT\frac{dn}{dt} \tag{S4}$$

The relation between pressure and deflection is described by the following equation.

$$P(\delta) = \frac{EtK(v)}{a^4}\delta^3 + \frac{4S_0}{a^2}\delta + P_{atm} = p_3\delta^3 + p_1\delta + P_{atm} \tag{S5}$$

where $S_0$ is the initial surface tension of the graphene which has a well-known value of 0.1 N/m[2,3]. The mechanical constants used are well established from numerous experiments of suspended graphene in a similar geometry[1,3-5]. For convenience, the parameters have been lumped into the constants $p_1$ and $p_3$. The relation between volume and deflection is described by the following equation.

$$V(\delta) = C(v)a^2\pi\delta + V_0 = v_1\delta + V_0 \tag{S6}$$

The constants for linear deflection term have been lumped together into $v_1$ for convenience.

The permeance is calculated by dividing dn/dt by the pressure difference of the effusing gas species. Classical effusion results in a linear relation between the rate of transport and the pressure difference, and therefore we define a constant value for normalized dn/dt, represented by $k$. The gas within the microcavity is assumed to be pure, and therefore total pressure $P(\delta)$ is equal to the partial pressure of the gas. $P_{ext}$ is the partial pressure of the gas species in atmosphere and is approximately zero for all of the gases tested (except $O_2$ and $N_2$).

$$\frac{dn}{dt} = k(P(\delta) - P_{ext}) \tag{S7}$$

By assuming this form for dn/dt, the differential equation can be written in terms of the deflection as follows.

$$\frac{d\delta}{dt}\frac{d(P(\delta)V(\delta))}{d\delta} = RTk(P(\delta) - P_{ext}) \tag{S8}$$

$$\frac{d\delta}{dt}\left((P_{atm}v_1 + p_1V_0) + (2p_1v_1)\delta + (3p_3V_0)\delta^2 + (4p_3v_1)\delta^3\right) = RTk(P_{atm} - P_{ext} + p_1\delta + p_3\delta^3) \tag{S9}$$

The dependence on the deflection is represented as a single arbitrary function, y.

$$y(\delta) = \frac{(P_{atm}v_1 + p_1V_0) + (2p_1v_1)\delta + (3p_3V_0)\delta^2 + (4p_3v_1)\delta^3}{(P_{atm} - P_{ext}) + p_1\delta + p_3\delta^3} \tag{S10}$$

$$y(\delta)\frac{d\delta}{dt} = RTk \tag{S11}$$

The differential equation is separable and can be solved.

$$\int_{\delta_0}^{\delta} y(\delta)d\delta = RTk\int_0^t dt \tag{S12}$$

$$Y(\delta) - Y(\delta_0) = RTkt \tag{S13}$$

The resulting form is a line. The values of the integral, $Y(\delta)$, can be calculated numerically for each experimental deflection point. A segment of these values is fit using the analytical least squares line fit, and gives results for $Y(\delta_0)$ and $k$ when temperature is known.

For the results of permeance or flux in the main paper, the fitting method is applied to a segment of 5 data points, and the resulting values are assigned to the centre data point. A value of k, the permeance, is calculated at each point by proceeding through the data set in this manner. Permeance values in the main text only include points with deflection above 50 nm. Points below 50 nm were excluded because small errors in the pressure correlation, and small amounts of air in the microcavity in some runs results in large errors in the permeance calculation at points lower than 50 nm deflection. When a full five points aren't available for the points at the beginning and end of the deflection data set, only the three or four nearest points are used to fit a value of permeation.

3. Deflection fitting (Fig.1s)

Figure 1s of the main paper plots an extrapolated fit of the deflection data before the sudden drop in deflection. This extrapolation represents the expected trajectory if the pore had continued in the state (blocked) at the start of the measurement. It was calculated by numerically linearizing the deflection data prior to the sudden drop (0 to 12 min) via S12 to the form of S13. From the analytical least squares fit, values for the two parameters, $Y(\delta_0)$ and $k$, can be determined based on all the points in the range of 0 to 12 min.

To create a fit and extrapolation for the deflection, a set of arbitrary, evenly spaced deflection values between the maximum and minimum deflections are put in the form of S13. The fitted values of the two parameters, $Y(\delta_0)$ and $k$, are used to solve for time at each of the arbitrary deflection points. Then the arbitrary deflection points are plotted against the calculated times as the fit and extrapolation.

4. Reversible switching of permeance from laser heating

Figure 2 of the main text shows switching of the permeance by laser induced heating which moves the AuNCs towards the pore site. This process is reversible. Further experiments on the same PSLG-AuNCs as in Figure 2 continue to show a slow permeance (gray coloured bar). Additional laser induced heating of the graphene resulted in a faster permeance shown by the magenta coloured bars in Supplementary Fig. 2.

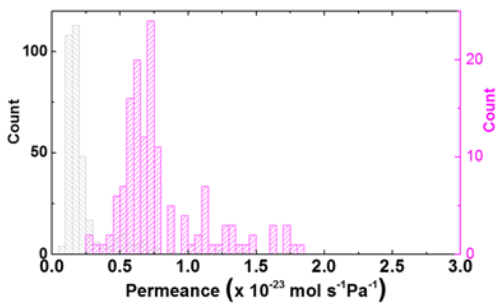

**Supplementary Fig. 2 Reversible switching of permeance from laser heating** Histogram of the permeance from (data in gray) to (data in magenta) by a second laser induced heating event.

5. Comparison to Classical Effusion

Supplementary Fig. 3 is used to compare selectivities predicted by the classical effusion model. Classical effusion requires the pore size to be smaller than the mean free path of the gas molecules which is around 60 nm at room temperature. However, we are in a regime where the angstrom-sized pore is much smaller than the mean free path and comparable to the size of the gas molecules. In this case, the molecular size, geometry, and chemistry should be considered.

In simple classical effusion, the permeance through a pore would be inversely proportional to the square root of the molecular mass ($M_w^{1/2}$), and the selectivity is the ratio of the $M_w^{1/2}$. This is 1.4, 3.2, 3.7, 4, 4.5, 4.7, 4.7 for $H_2$ to He, Ne, $N_2$, $O_2$, Ar, $N_2O$, $CO_2$, respectively. However, the selectivity from our experiment is dramatically different as predicted from classical effusion (Table 1). This is most clearly seen when comparing $O_2$ and $N_2$. The permeance of $O_2$ is 2.6x the permeance of $N_2$ despite $O_2$ having a larger $M_w$ suggesting that our experiment is not in the classical effusion regime and we must consider the kinetic diameter. $O_2$ has a slightly smaller kinetic diameter compared to $N_2$. Supplementary Fig. 3d which plots the permeance of the Noble gases measured show strong deviations from classical effusion. The permeance of Ar is significantly lower than expected for classical effusion suggesting that molecular sieving is taking place based on the kinetic diameter.

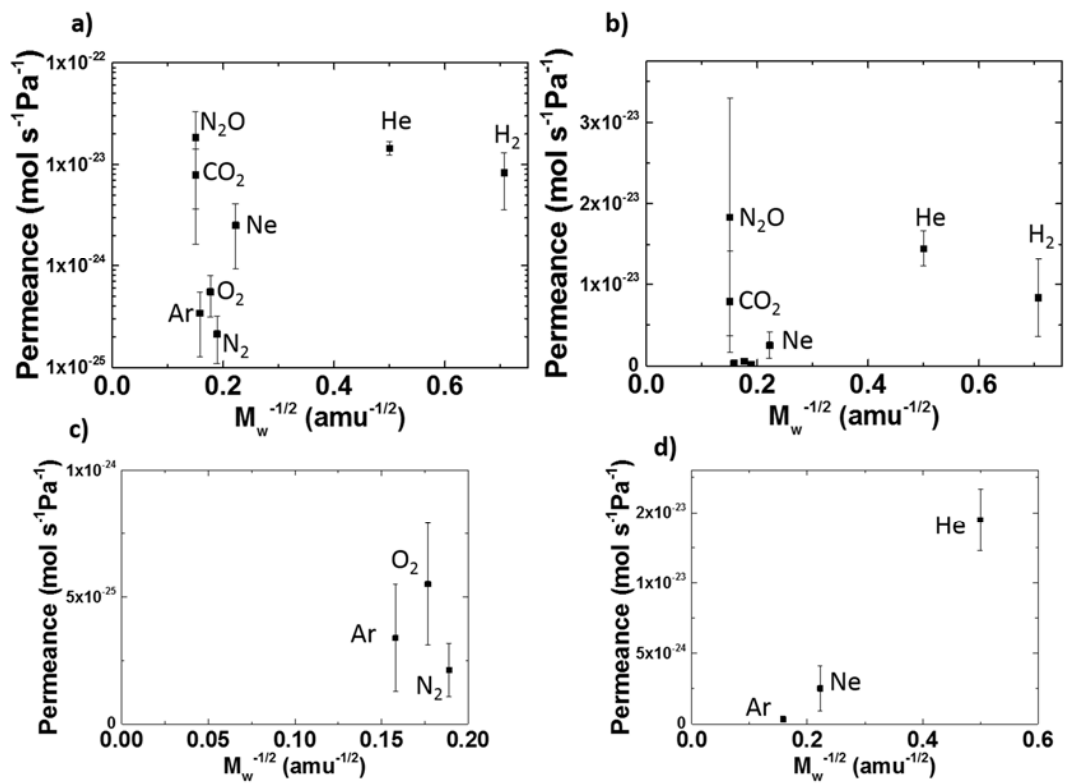

**Supplementary Fig. 3 Comparison of the permeance to classical effusion.** (a) Permeance for the same sample as Figure 3 & 4 plotted versus the inverse square root of the molecular mass of He, Ne, Ar, $H_2$, $N_2O$, $CO_2$, $O_2$, and $N_2$ in log scale. (b) (a) in linear scale. (c) Permeance versus square root of the molecular mass of Ar, $O_2$, and $N_2$. (d) Permeance versus square root of the molecular mass of the noble gases. Error bars represent ±s.d. for different measurements on the same membrane.

TABLE 1 Selectivities of $H_2$ to other gases

|  | He | Ne | $N_2$ | $O_2$ | Ar | $N_2O$ | $CO_2$ |
|---|---|---|---|---|---|---|---|
| Selectivity predicted by classical effusion | 1.4 | 3.2 | 3.7 | 4 | 4.5 | 4.7 | 4.7 |
| Experimental selectivity | 0.6 | 3.3 | 39.2 | 15.1 | 24.6 | 0.5 | 1.1 |

6. Maximum deflection vs. time for the PSLG in Figure 3, 4, and Supplementary Fig. 3

The max deflection vs. time for He, Ne, Ar, $H_2$, $N_2O$, $CO_2$, $O_2$, and $N_2$ are shown in Supplementary Fig. 4. Different colours represent different measurements for the same

PSLG. From this data, we extracted the leak rate dn/dt vs. pressure difference $\Delta p$ and the corresponding permeance. The permeance after etching for $O_2$ is 5.5 x $10^{-25}$ mol-$s^{-1}$-$Pa^{-1}$, and the permeance for $N_2$ is 2.1 x $10^{-25}$ mol-$s^{-1}$-$Pa^{-1}$.

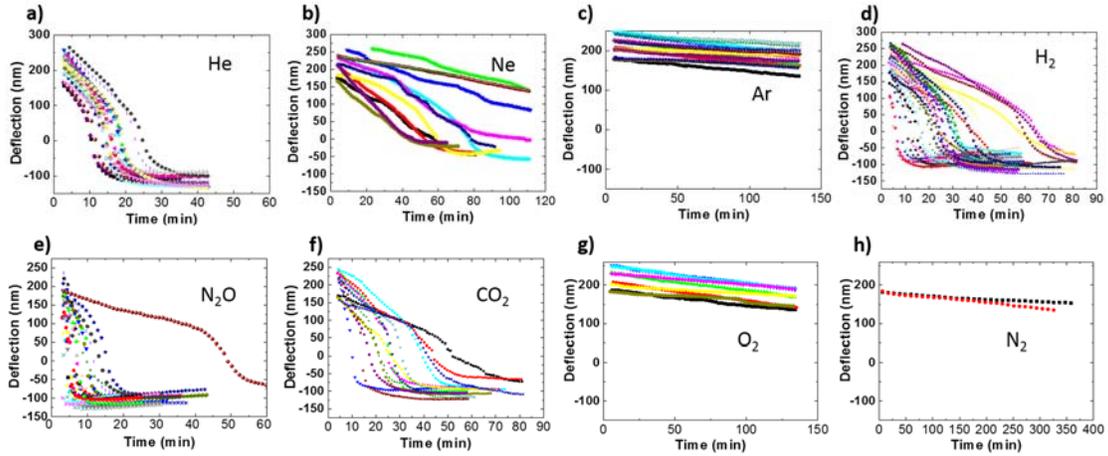

**Supplementary Fig. 4 Additional gas permeation data.** Maximum deflection vs time of multiple gases for the PSLG in Fig. 3, 4, and Supplementary Fig. 3.

7. Hidden Markov model

For fitting permeance versus time to discrete states, we applied a hidden Markov model (HMM). HMM describes a system that can switch between various states, however the states are not directly observable and must be inferred from changes in output. This type of modelling has been applied in fluorescence and sensing applications to model the changes in observed fluorescence from interactions with an analyte[6]. The states and fit were calculated using the program HaMMy, which was originally developed for HMM analysis of FRET systems[7].

The data from multiple runs was analysed together by concatenating all data sets so that they have a shared time axis, called "observed time". This represents the time in which data was being measured. The experiments for Ne were done over the course of five days, and the time between experiments is excluded for the fitting but noted by dashed lines in Figure 4. For estimating the frequency of transitions between states, the transitions that occurred between the end of the previous experiment and the start of the next are excluded. Only points and time when the deflection is above 50 nm are included. The fitting algorithm fits the data to states that are distinct from each other and treats the transitions between states as instantaneous. The program was used to fit up to ten states, but it can also return empty states if the number of recognizable states is lower than ten.

8. Markov network for multiple two-state pores

We investigated how well the observed data fit to the expected Markov network for multiple two-state pores as compared to a single pore with many states. In the case of multiple two-state pores, the properties of each observed state are constrained and set by the individual properties of isolated, independent pores. In comparison, a system with one pore having many states does not inherently require there to be any relation amongst the observed states.

We examine how well the permeance values of the observed states fit to a 3-pore system by applying least squares to fit the data to eight total states with four variable parameters. The values of these eight states are set by the combinations of the two states, high and low, of each of the three pores, given by equation S14; $x$ is the combined permeance of the low states for all three pores; $y_a$, $y_b$, and $y_c$ are the difference between the high and low states for the first, second, and third pores respectively; and $\Pi_i$ is the permeance value for the ith observable state. Note that the states are not necessarily ordered in terms of increasing permeance value.

$$\begin{pmatrix} y_a \\ y_b \\ y_c \end{pmatrix} \begin{pmatrix} 0 & 0 & 0 \\ 0 & 1 & 0 \\ 0 & 0 & 1 \\ 0 & 1 & 1 \\ 1 & 0 & 0 \\ 1 & 1 & 0 \\ 1 & 0 & 1 \\ 1 & 1 & 1 \end{pmatrix} + x = \begin{pmatrix} \Pi_1 \\ \Pi_2 \\ \Pi_3 \\ \Pi_4 \\ \Pi_5 \\ \Pi_6 \\ \Pi_7 \\ \Pi_8 \end{pmatrix}$$

(S14)

Using equation S14 to calculate states based on the four parameters from the three two-state pores, we used least squares to fit the experimental data points to discrete states. For the data set collected using Ne gas, this yielded $\{x, y_a, y_b, y_c\} = \{1.09, 0.83, 1.81, 4.04\} *10^{-24}$ mol m$^{-2}$ Pa. The fit to the data set and comparison of the state values to the fit generated with the HaMMy program are plotted in Supplementary Fig. 5; this figure is nearly identical to Fig. 4 in the main manuscript, with the difference that the applied fit is the least squares (LS) fit constrained by the relations of the three pore Markov network rather than the fit using the HaMMy program.

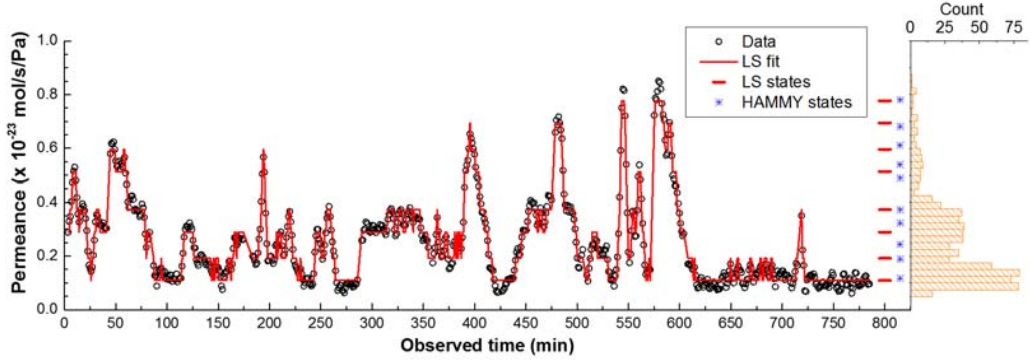

**Supplementary Fig. 5: Plot of the least squares (LS) fit of the permeance data to the interrelated eight states of the three two-state pore model.** A summary of the state values and comparison to previous fitting with HaMMy along the right

The fraction of time spent in each of the observable states should also be interrelated in a 3-pore system. The relations are described by equations S15-S22, where $P_i$ is the fraction of time spent (or probability of finding the system) in the ith observable state and $p_a$, $p_b$, and $p_c$ are the fraction of time each of the first, second, and third pores each respectively spend in their high permeance state.

$$P_1 = P_{000} = (1-p_a)(1-p_b)(1-p_c) \tag{S15}$$

$$P_2 = P_{010} = (1-p_a)p_b(1-p_c) \tag{S16}$$

$$P_3 = P_{001} = (1-p_a)(1-p_b)p_c \tag{S17}$$

$$P_4 = P_{011} = (1-p_a)p_b p_c \tag{S18}$$

$$P_5 = P_{100} = p_a(1-p_b)(1-p_c) \tag{S19}$$

$$P_6 = P_{110} = p_a p_b(1-p_c) \tag{S20}$$

$$P_7 = P_{101} = p_a(1-p_b)p_c \tag{S21}$$

$$P_8 = P_{111} = p_a p_b p_c \tag{S22}$$

In an exact system, any set of three from equations S15-S22 can be used to extract values for underlying parameters $p_a$, $p_b$, and $p_c$. The analytical solutions in these cases contains a square root, and the contents of the square root can be used as a quick check of whether there is a real solution or not as a an indicator of whether the system can be described by the three two-state pore model. Two sets of these expressions are given in equations S23 and S24, which use the sets of equations {S15, S21, S22} and {S15, S16, S22} respectively. The values $r_1$ and $r_2$ are terms that must be greater than zero for there to be a real solution, and the $P_i$'s can be extracted as observables from the fit to the eight states.

$$r_1 = -4P_7^2(P_7+P_8)+(P_7(1-P_1+P_8)-P_1P_8+P_7^2)^2 \tag{S23}$$

$$r_2 = -4P_2P_8(P_1+P_2)+(P_2(P_1+P_2-1)-(P_1+P_2)P_8)^2 \tag{S24}$$

Calculation of these values with the data set collected using Ne gas yields $r_1=2*10^{-5}$ and $r_2=3*10^{-3}$; the positive values show that a real solution is possible based on those observed values.

The two sets of equations, equation S14 and equations S15-S22, describe a result in an over specified system where the number of observable values and their equations is greater than the variables. Hence the fidelity of the observed data to the model of three two-state pores is best described in terms of a fit of the equations to their observed values in the experimental data. Rather than selecting a subset of the equations, the three underlying values of $p_a$, $p_b$, and $p_c$ were fit with least squares between the calculated $P_i$ based on equations S15-S22 and the observed state times.

The fitted values of $p_a$, $p_b$, and $p_c$ were found to be 0.381, 0.376, and 0.110 in order of increasing $y_i$ value. The comparison of the eight $P_i$ values calculated with those fitted parameters to the observed values is represented in Supplementary Fig. 6.

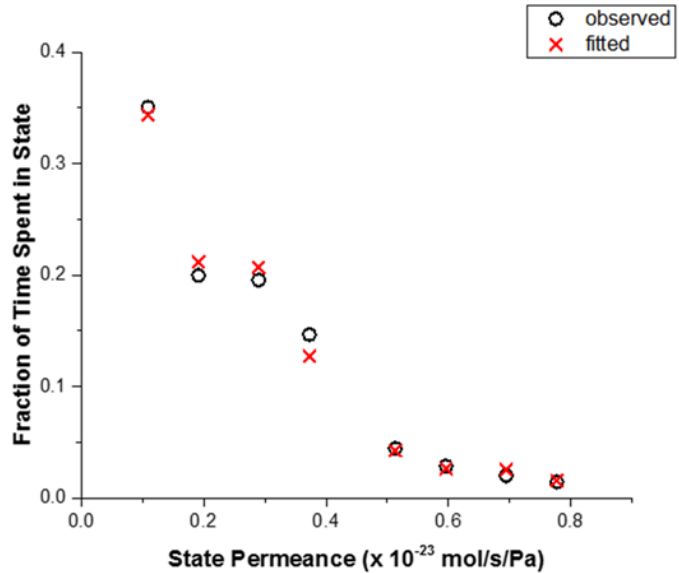

**Supplementary Fig. 6** Comparison of the values for fraction time spent, Pi, between the observed values from the experimental results and the values calculated from the three fitted parameters of the three two-state pore model.

The observed dwell times and corresponding fraction of time spent being fit well shows that the experimental membrane is consistent with a three pore model. Furthermore, we confirmed that similar simulated data with eight states, unconstrained by the three pore model, could not be fit as well as the experimental results.

9. Transition activation energy estimation

The frequency of transitions between states can be used to estimate the activation energy of the pore transitions. For Ne, the hidden Markov modelling identified 54 transitions that did not occur during the time between experimental runs. The total observed time was 785 minutes. Therefore, the frequency is about 1 transition per 15 min, or $1.1 \times 10^{-3}$ 1/s; with the assumption there are three active pores contributing to the fluctuations, the average frequency for an individual pore should be a factor of three less, $3.7 \times 10^{-4}$ 1/s. The kinetic rate constant is approximately equivalent to the frequency. For an elementary process, the kinetic rate constant consists of an attempt frequency, $A$, and an exponential dependence on activation energy.

$$k = Ae^{-E_a/RT} \approx 3.7 \times 10^{-4} \, 1/s \qquad (S25)$$

Molecular vibrations typically occur within a few orders of magnitude of $10^{13}$ 1/s[8]. Assuming a temperature of 298 K, the average activation energy for a transition, $E_a$, can be estimated as 1.0 eV.

10. Estimation of change in permeance due to pore rearrangement

We attribute the stochastic switching in permeance to small fluctuations at the pore site which affects the transmission probability. One possible source of fluctuations is reaction of the pore edge functionalization with atmospheric species while another is pore isomerization, in which the pore's bonds transition between stable chemical states through rearrangement, but no gain or loss of the atoms around the pore. Supplementary Fig. 7a illustrates a simple possible rearrangement: moving a single carbon atom in a model pore which excludes functionalization. The effective size is approximately the expected pore size based on the observed molecular sieving. Even for this simplified model, the small rearrangement results in a relatively significant change in the energy barrier, as shown in Supplementary Fig. 7b. That change in the model pore's estimated energy barrier gives an expected change in the permeance of around a factor of two or three, which is consistent with the magnitude of the experimentally observed changes in Fig. 4. The sensitivity of the permeance to small rearrangements at a single pore further supports the hypothesis that a single pore (or a very small number of pores) is responsible for the observed gas transport in our measured porous graphene.

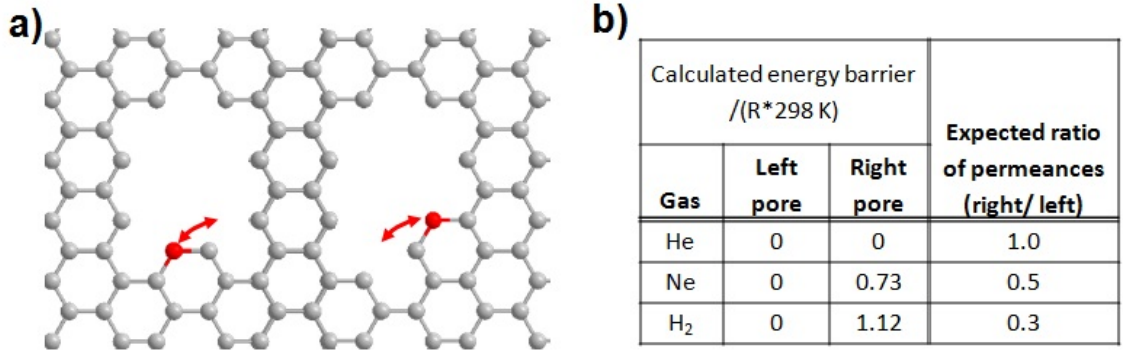

**Supplementary Fig. 7 Atom rearrangement at pore mouth.** (a) Representation of a simple model pore rearrangement. (b) Estimation of barrier energy for two pores depicted in (a) and the expected ratio of permeance between the two.

In Supplementary Fig. 7b, it is shown that a small rearrangement in the pore could result in a meaningful change in the barrier energy and the expected permeance. The ratio is derived from an alternate form of equation 2 in the main text rearranged in terms of permeance, $\theta$. The transmission coefficient, $\gamma$, is expanded into a geometric area term, $a_g$, and a barrier energy, $E_a$, term.

$$\theta = \frac{1}{\Delta p}\frac{dn}{dt} = \frac{\gamma}{\sqrt{2\pi M_w RT}} = \frac{a_g e^{-E_a/RT}}{\sqrt{2\pi M_w RT}} \tag{S26}$$

When there is a pore rearrangement, both component terms of the effective pore area can change. However, as a first order approximation, we will assume that the geometric area term remains approximately constant. Therefore, the ratio of the permeances reported in Supplementary Fig. 4b is simply the following:

$$\frac{\theta_2}{\theta_1} = \frac{a_{g,2} e^{-E_{a,2}/RT}}{a_{g,1} e^{-E_{a,1}/RT}} \sim \frac{e^{-E_{a,2}/RT}}{e^{-E_{a,1}/RT}} \tag{S27}$$

The energy barriers for pore configuration 1, $E_{a,1}$, and pore configuration 2, $E_{a,2}$, are calculated using a single-centre Lennard-Jones potential with the following parameters.

TABLE 2 Parameters for Lennard-Jones potential calculation

| Center | σ (Å) | ε / $k_B$ (K) |
|---|---|---|
| C[9] | 2.960 | 34.2 |
| Ar[10] | 3.542 | 93.3 |
| He[10] | 2.551 | 10.22 |
| Ne[10] | 2.820 | 32.8 |
| $CO_2$[10] | 3.941 | 195.2 |
| $H_2$[10] | 2.827 | 59.7 |
| $N_2$[10] | 3.798 | 71.4 |
| $N_2O$[10] | 3.828 | 232.4 |
| $O_2$[10] | 3.467 | 106.7 |

Supplementary Fig. 7b only includes He, $H_2$, and Ne. The full list of energy barriers is listed below. Energy barriers that are negative mean that the interaction energy for being in the centre of the pore is more favourable than as a free gas molecule; the energy barrier is treated as zero in this case. For the larger molecules the energy barrier is larger and means those molecules would permeate very slowly, effectively blocked relative to the lower barrier. $CO_2$ and $N_2O$ show large energy barriers even though experiments show they permeate quickly; as discussed in the manuscript, this is believed to be the result of favourable electronic interactions that lower the total interaction energy, which is not captured by the simple Lennard-Jones (L-J) potential.

Table 3 Calculated Energy Barriers

| Gas | L-J Energy barrier $E_a/(R \times 298)$ | |
|---|---|---|
| | Supplementary Fig. 7a Left | Supplementary Fig. 7a Right |
| He | 0 (-0.83) | 0 (-0.37) |
| Ne | 0 (-0.96) | 0.73 |
| $H_2$ | 0 (-1.36) | 1.12 |
| $O_2$ | 7.29 | 23.8 |
| Ar | 8.55 | 25.7 |
| $N_2$ | 14.5 | 36.8 |
| $CO_2$ | 45.5 | 109.2 |
| $N_2O$ | 40.9 | 102.4 |

11. Movement of gold clusters by AFM tip

Supplementary Fig. 8 shows the movement of Au clusters on the device in Figure 1b- 1d after the $N_2$ gas has leaked out. One can see Au clusters being pushed around presumably due to interactions with the AFM tip (see the slanted lines in the 2$^{nd}$ and 4$^{th}$ image of Supplementary Fig. 8).

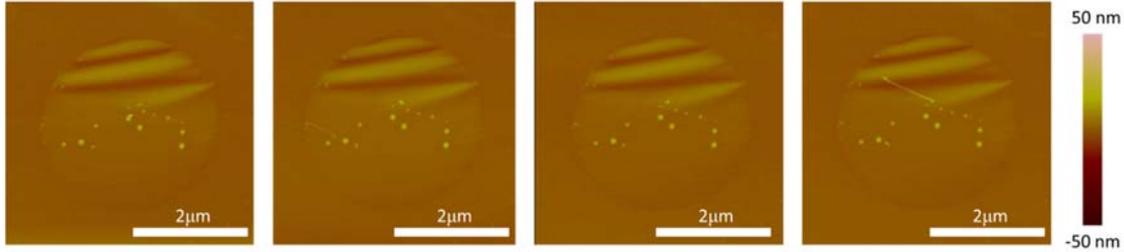

**Supplementary Fig. 8 Movement of Au clusters on suspended graphene with one pore at the centre introduced by the voltage pulse method.** (a-d) Sequential AFM scans showing the movement of the Au clusters.

12. Additional sample showing laser induced changes to the permeance

The device in Figure 1b -1d (main text) was heated in a laser in a similar manner to the device in Figure 2 (main text). These 2 devices had their pore fabricated in a very different manner – UV etching vs. voltage pulse method, yet the results are similar. Supplementary Fig. 9 shows movement of gold clusters before and after laser heating.

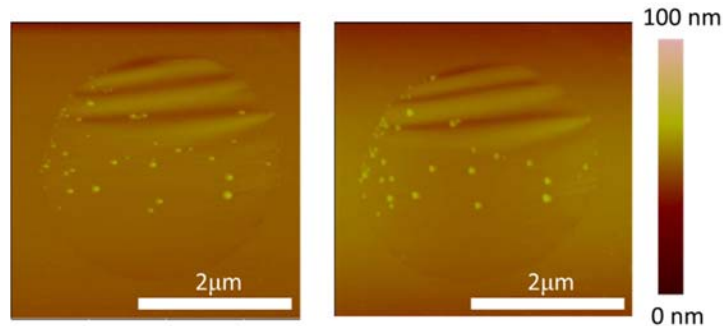

**Supplementary Fig. 9 Movement of Au clusters by laser heating.** The configuration of Au clusters changes before (a) and after (b) laser heating.

The permeance of this device was measured before and after laser heating in a manner similar to that of Figure 2 (main text). Supplementary Fig. 10 shows the maximum deflection vs. time for the device in Figures 1b-1d and Supplementary Fig. 8-9. The green and dark blue points are prior to laser heating. The difference might have resulted from a "random" change in the gold particle configuration induced by interactions with the AFM tip. The behaviour observed is similar to what was observed in Figure 1k of the main text.

After laser heating, the pore "closed" and demonstrated a much lower leak rate than before heating. This is consistent and similar to what was observed in Figure 2 (main text). The device was damaged after this measurement and therefore no further switching was possible like what was seen in Figure 2 (main text) and Supplementary Fig. 2. The similarity between the results on the single pore fabricated by the voltage pulse method and the single pore fabricated by UV etching demonstrates the robustness of the measurement technique to observe and control gas transport through single pores in graphene molecular valves.

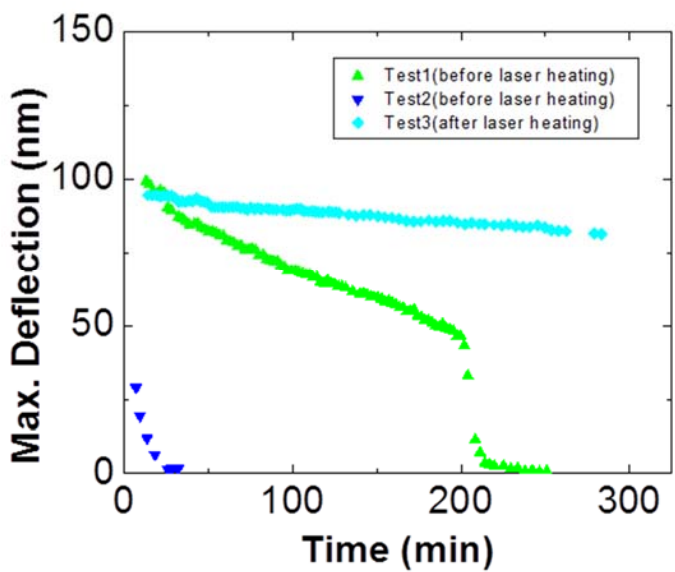

**Supplementary Fig. 10 Laser induced heating to change the leak rate of $N_2$ across a porous graphene valve fabricated by the voltage pulse method.** Maximum deflection vs. time for 3 different measurements on the porous graphene valve from Figure 1b-d in the main text. Test 1 and Test 2 were performed before laser heating while Test 3 was done after laser heating.

**Supplementary References:**